\documentclass[sigconf, balance = false]{acmart}

\usepackage{booktabs} %

\acmJournal{TOG}

\copyrightyear{2023} 
\acmYear{2023} 
\setcopyright{acmlicensed}
\acmConference[SIGGRAPH '23 Conference Proceedings]{Special Interest Group on Computer Graphics and Interactive Techniques Conference Conference Proceedings}{August 6--10, 2023}{Los Angeles, CA, USA}
\acmBooktitle{Special Interest Group on Computer Graphics and Interactive Techniques Conference Conference Proceedings (SIGGRAPH '23 Conference Proceedings), August 6--10, 2023, Los Angeles, CA, USA}
\acmPrice{15.00}
\acmDOI{10.1145/3588432.3591501}
\acmISBN{979-8-4007-0159-7/23/08}

\usepackage{hyperref}
\usepackage{url}
\usepackage{wrapfig}
\usepackage{xcolor}
\usepackage{colortbl}
\usepackage{multirow}

\usepackage{wrapfig}
\usepackage{nicefrac}
\usepackage{mathtools}
\usepackage{graphicx}
\usepackage{booktabs} 
\usepackage{combelow} %
\usepackage{tabularx} %
\usepackage{colortbl} %

\usepackage{manfnt} %

\usepackage{caption}
\usepackage{subcaption}

\usepackage{amsmath}
\usepackage{amsfonts}
\usepackage{amsbsy}

\usepackage{cleveref}

\usepackage{algpseudocode}

\usepackage[ruled]{algorithm2e} %

\SetAlFnt{\small}
\SetAlCapFnt{\small}
\SetAlCapNameFnt{\small}
\SetAlCapHSkip{0pt}

\usepackage[utf8x]{inputenc}

\usepackage{xr}

\usepackage{pdfpages}
\def\tlmV{{\widetilde{\matFont{V}}}}
\newcommand{\tl}[1]{\widetilde{#1}}
\DeclareMathOperator{\Tr}{Tr}

\DeclareMathOperator*{\argmin}{arg\,min}
\DeclareMathOperator*{\minimize}{minimize}

\newcommand{\refequ}[1] {Eq.~\ref{equ:#1}}

\newcommand{\reffig}[1] {Fig.~\ref{fig:#1}}

\newcommand{\refsec}[1] {Sec.~\ref{sec:#1}}

\newcommand{\refalg}[1] {Alg.~\ref{alg:#1}}

\newcommand{\st}{\text{s.t.}}

\newcommand{\capFont}[1]{\mathcal{#1}}

\def\MN{{\capFont{N}}}

\def\MS{{\capFont{S}}}

\newcommand{\vecFont}[1]{\mathbf{#1}}

\def\vb{{\vecFont{b}}}

\def\vd{{\vecFont{d}}}

\def\vp{{\vecFont{p}}}

\def\vx{{\vecFont{x}}}

\newcommand{\matFont}[1]{\mathbf{#1}}
\def\mA{{\matFont{A}}}
\def\mB{{\matFont{B}}}

\def\mD{{\matFont{D}}}

\def\mF{{\matFont{F}}}

\def\mI{{\matFont{I}}}

\def\mL{{\matFont{L}}}
\def\mM{{\matFont{M}}}

\def\mR{{\matFont{R}}}
\def\mS{{\matFont{S}}}
\def\mT{{\matFont{T}}}
\def\mU{{\matFont{U}}}
\def\mV{{\matFont{V}}}
\def\mW{{\matFont{W}}}
\def\mX{{\matFont{X}}}

\def\mZ{{\matFont{Z}}}

\newcommand{\gray}[1]{{\color{gray}#1}}

\newcommand{\changed}[1]{{#1}}
\colorlet{RED}{red} %

\newcommand{\sym}[1]{\text{sym}(#1)}

\newcommand{\lossname}{\changed{\text{SC-L1} loss}}

\newcommand{\lossnameshort}{\changed{\text{SC-L1}}}
\newcommand{\Lossnameshort}{\changed{\text{SC-L1}}}
\newcommand{\mlossname}{\text{\changed{SC-L1}}}

\newlength\savedwidth

\begin{document}

\title{Local Deformation for Interactive Shape Editing}

\author{Honglin Chen}
\affiliation{%
  \institution{Columbia University}
  \city{New York City}
  \state{NY}
  \country{USA}
  }
\email{honglin.chen@columbia.edu}

\author{Changxi Zheng}
  \affiliation{%
    \institution{Columbia University}
    \city{New York City}
    \state{NY}
    \country{USA}}  
  \email{cxz@cs.columbia.edu}

\author{Kevin Wampler}
  \affiliation{%
    \institution{Adobe Research}
    \city{Seattle}
    \state{WA}
    \country{USA}}
  \email{kwampler@adobe.com}

\begin{abstract}
We introduce a novel regularization for localizing an elastic-energy-driven deformation to only those regions being manipulated by the user.
Our local deformation features a natural region of influence, which is automatically adaptive to the geometry of the shape, the size of the deformation and the elastic energy in use.
We further propose a three-block ADMM-based optimization to efficiently minimize the energy and achieve interactive frame rates. 
Our approach avoids the artifacts of other alternative methods, is simple and easy to implement, does not require tedious control primitive setup and generalizes across different dimensions and elastic energies.
We demonstrates the effectiveness and efficiency of our localized deformation tool through a variety of local editing scenarios, including 1D, 2D, 3D elasticity and cloth deformation. 
\end{abstract} 


\begin{teaserfigure}
\centering
  \includegraphics[width=1.0\linewidth]{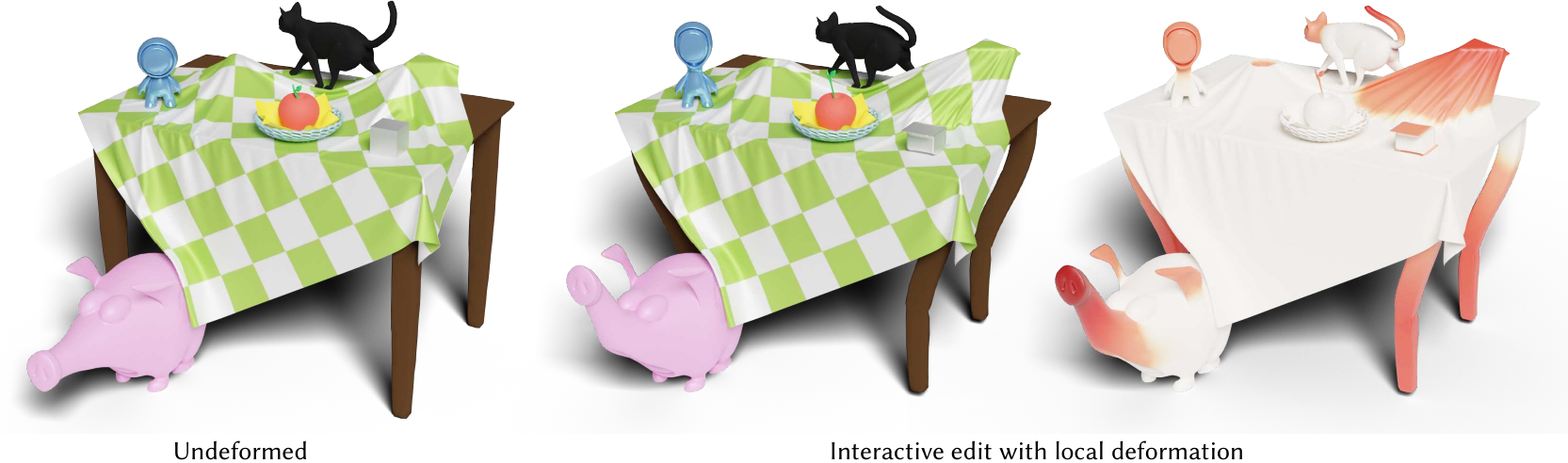}
  \caption{\changed{Our method enables the user to edit shapes
      in an interactive and physically plausible way.
      The edit is \emph{local}, meaning that the user can focus on
      one region of the complex scene without worrying about
      inadvertent changes elsewhere.
      To visualize the locality, in the rightmost figure
      we highlight the regions where the vertex
      displacement is larger than $10^{-3}$ in red.
(Undeformed scene shapes thanks to \cite{zhang2022mgCloth})}
  }
  \label{fig:teaser}
\end{teaserfigure}

\maketitle

\section{Introduction}

Local deformation is a core component in modeling and animation.
In a localized deformation, only the parts of the shape near where the user is currently manipulating move\textemdash everything else stays still, ensuring that the user can focus entirely on one region of the shape without worrying about inadvertent changes elsewhere.
However, existing localized deformation tools tend to have practical impediments for interactive design: they are either too slow to run, unaware of the geometry, introduce artifacts, or require a careful control point setup.

When global deformation is acceptable, a widely useful approach is to solve for the deformation by minimizing an elastic energy defined over the shape, subject to positional constraints derived from the user's input.
This paradigm has many advantages.
The deformation accounts for the geometry of the shape, generalizes well to 2D, 3D, and cloth, and the elastic energy can be used to model a wide range of both real-world and stylized materials.
Unfortunately elastic energy minimization is by its nature global, and jointly solves for all the degrees of freedom in the shape.
This necessitates a rigging step to ``pin down'' certain aspects of the deformation lest the optimizer move them.
This limits the applicability of such methods to situations where a suitable rig is available, or the region of influence for a deformation is known in advance.

We seek to combine the advantages of elastic energy minimization with the locality of sculpting-style tools.
In doing so, we also want the locality of a deformation to be \emph{automatic}, \emph{natural} and \emph{efficient}.
The locality of an edit should be automatic in that the region of influence (ROI) of the deformation should scale automatically depending on the size of the desired deformation.
In addition, although we do not require a rig, in cases where a rig or other constraints have been placed on the shape, the ROI needs to automatically adapt to them.
We further want the notion of locality to be \emph{natural} in that it adjusts to both the geometry of the shape and the elastic energy driving its deformation, where changes to either will lead to a fitting change in the ROI.
Finally, we require that the method be fast enough to run in real time.

We achieve this with the following contributions:
\begin{itemize}
	\item We introduce a novel \changed{deformation} regularizer, called a \changed{smoothly clamped $\ell_1$ (\lossnameshort{}) loss} which augments an elastic energy with a notion of locality.  \Lossnameshort{} regularization is simple to implement, and avoids artifacts of previous methods.
	\item We enable real-time localized deformation with an ADMM-based optimization algorithm for \lossnameshort{}-regularized deformation which is significantly faster than prior work using a group lasso regularizer.
\end{itemize}
We illustrate the utility of \lossnameshort{} regularization on a wide range of examples, including multiple different elastic energies, 1D curves, 2D and 3D meshes, and cloth.
This provides a localized deformation tool which avoids artifacts of other regularizers, is easy to implement, generalizes across different dimensions and material models, and performs fast enough to run in real time.

\vspace{-2mm}
\section{Related Work}\label{sec:related}

Shape deformation algorithms in computer graphics have typically fallen into one of two categories, which we will call the \emph{direct} approach and the \emph{optimization} approach.
In the direct approach, the shape's deformation is an explicit function of the user's input, often either by modifying some high-level parameterization or by applying a pre-specified deformation field.
In the optimization approach, the shape deformation is an indirect product of the user's input combined with an elastic energy, and the deformation itself is only known after an optimization process has converged to minimize this energy.
These two approaches differ in how (and often whether) they enforce the locality of a deformation.

\vspace{-2mm}
\subsection{Locality in High-Level Parameterizations}

A time-honored and common approach for localized deformation is the direct manipulation of high-level parameters.
From the onset, a notion of locality is baked into these parameters, which can directly encode the shape itself, often using splines \cite{10.5555/174506} or a ``rig'' to control a shape's deformation, as with cage-based generalized barycentric coordinates \cite{joshi2007harmonic,lipman2008green}, linear-blend skinning \cite{Magnenat1989LBS,jacobson2011BBW,wang2015bhc}, lattice deformers \cite{Sederberg1986lattice, Coquillart1990lattice}, wire curves \cite{singh1998wire}, or learned skinning weights \cite{genova2020local}, to
list a few.

Approaches of this nature, although popular in computer graphics, have a few disadvantages.
Firstly, since the locality is baked into the parameterization, it cannot easily adapt based on changes in the deformation.
Secondly, without an optimization step\changed{,} even in cases where a capable elastic energy is within reach there is no way to incorporate it into the deformation.

\begin{figure}[t]
  \centering
  \includegraphics[width=1.0\linewidth]{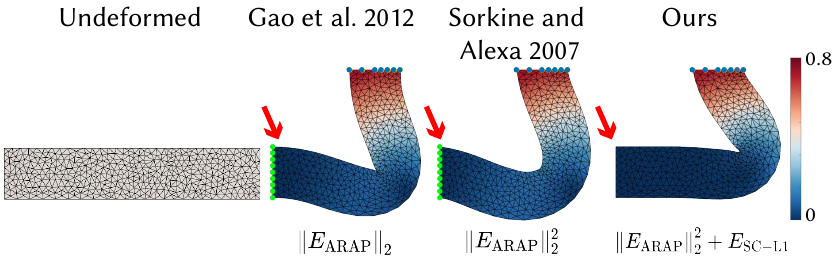}
  \caption{Directly applying different norms to an elastic energy 
      leads to global deformation, and thus the user needs to
      carefully set up additional fixed handles (green points) 
      to keep the shapes from freely moving around in the space. }
  \label{fig:bar_2d_Lp}
\end{figure}  

\vspace{-2mm}
\subsection{Locality in Deformation Fields}

Instead of providing localized deformation via pre-chosen parameters, it is also possible to define a localized deformation field, then apply it to a shape.
These methods often follow a ``sculpting'' metaphor, and include simple move, scale, pinch, and twist edits as well as more sophisticated operations \cite{cani2006survey}.
Recently, 
De Goes and James~\shortcite{DeGoes2017Kelvinlets} introduced \emph{regularized
Kelvinlets}, which provides real-time localized volumetric control based on the
regularized closed-form solutions of linear elasticity.
These closed-form solutions were later extended to handle dynamic secondary motions
\cite{DeGoes2018dynamicsKelvinlets}, sharp deformation
\cite{DeGoes2019sharpKelvinlets} and anisotropic elasticity
\cite{chen2022goGreen}.

These approaches allow for local deformation with real-time feedback. %
However, as they are designed for digital sculpting, these methods usually require the user to explicitly pick the falloff of the brushes.
Furthermore, these methods are usually based on Euclidean distance, unaware of the shape's geometry.
In contrast, our method is shape-aware and enables automatic dynamic region of influence with interactive feedback.

\vspace{-2mm}
\subsection{Localized Optimization via an ROI}

The most natural formulation of optimization-based deformation editing is by
solving globally for the entire shape's deformation at once \changed{\cite{smith2019eigensystems,zhu2018bcqn,shtengel2017composite}}.
Nevertheless, there are methods attempting to enforce locality in the
shape optimization process.  Previous methods of this sort have often
computed the ROI of a manipulation as a preprocessing step, then restricting
the optimization to only move parts of the shape within this ROI.  
The ROI can be taken as an input~\cite{alexa2006laplace} or based
on a small amount of user markup~\cite{kho2005sketch,zimmermann2007sketch,luo2007roi}.  
Other methods combine deformation energies with handle-based systems, including skeleton rigs
\cite{jacobson2012ARAPskinning,hahn2012rig,kavan2012elasticityRig} and cages
\cite{ben-chen2009cage}.

Unfortunately, in many contexts, the ROI is hard or impossible to know in advance.
This is particularly the case when constraints are involved, or where it is not known in advance if a deformation
will be small (best fitting a small ROI) or large (best fitting a large ROI).
In addition, the correct ROI may also depend on the elastic energy driving the
deformation, thus difficult to account for when the ROI calculation is
decoupled in a separate step.
\begin{figure*}[tp]
  \centering
  \includegraphics[width=1.0\textwidth]{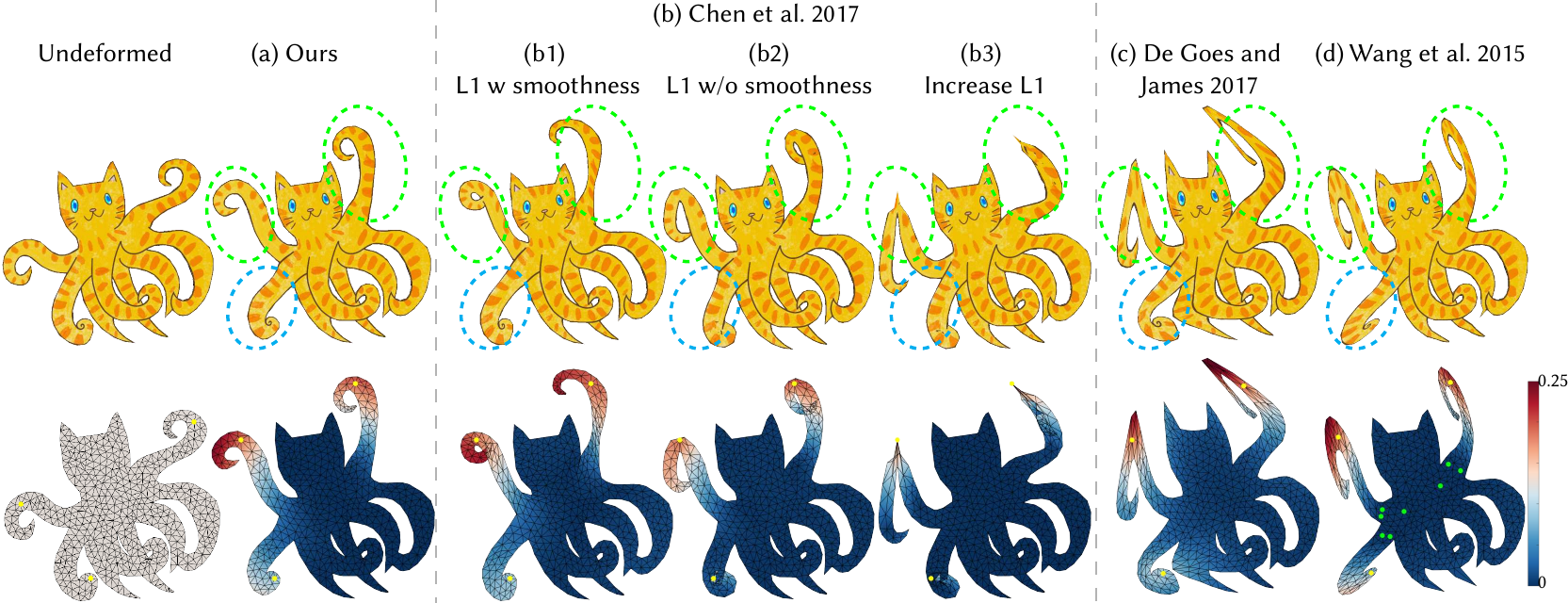}
  \caption{Our method (a) produces \emph{local} deformation which is smooth, shape-aware and has automatically adaptive region of influence without the need of additional handles. 
    Here we highlight all the handles that have been moved in yellow.
    For the other alternatives, $l_1$-norm-based method (b) contains artifacts due to the use of $l_1$ norm (see the green highlighted regions); regularized Kelvinlets technique (c) is based on Euclidean distance and unaware of the geometry (see the blue highlighted regions); and biharmonic coordinates approach (d) requires a careful placement of additional handles (green) to explicitly control the region of influence.
    \changed{Note that the use of $l_1$-norm (b) leads to either small global motions and minor artifacts (b2, smaller $l_1$-norm) or no global motions but significant artifacts (b3, larger $l_1$-norm).
    Thus \cite{chen2017L1ARAP} mitigates it with a smoothing regularizer (b1), but this changes the elastic energy and it no longer deforms like a localized ARAP.
    }
    }
  \label{fig:octocat_bird_vs_all}
  \vspace{-6pt}
\end{figure*}

\vspace{-2mm}
\subsection{Localized Optimization via Sparsity Norms}
\changed{
Sparsity-inducing norms, such as the smooth $\ell_0$ norm, have been widely applied to many domains and problems including medical image reconstruction \cite{xiang2022smooth_l0},
sparse component analysis \cite{mohimani2007smooth_l0} and UV mapping \cite{poranne2017autocuts}.
}
To allow an adaptive ROI while preserving the benefits of optimization-based
deformation, a few works have adopted a sparsity-inducing norm, typically 
a $\Vert x \Vert_1$ or $\Vert x \Vert_2$ norm, in their energy, 
which is then minimized by Alternating Direction Method of Multipliers
(ADMM) \changed{\cite{boyd2011ADMM,peng2018anderson,zhang2019admm}} or Augmented Lagrangian Method (ALM)
\cite{Bertsekas1996ALM}. 
We refer the readers to \cite{xu2015survey} for a
survey on sparsity in geometry modelling and processing. 

Several methods of this variety rely on sparsity-inducing norm formulated
as a sum of $\Vert x \Vert_2$ norms, referred to as $\ell_{2,1}$ or
$\ell_1/\ell_2$ norms, or a \emph{group lasso} penalty.  They have been applied
in a preprocessing phase to compute sparse deformation modes
for interactive local control~\cite{neumann2013sparseMode,deng2013local,brandt2017compressedMode}.
However, the deformation is limited by the
linear deformation modes and thus struggles with large deformation.

Another class of methods adds a sparsity-induced regularization to an elastic
energy optimization to achieve local deformation. Gao et al.~\shortcite{lin2012LpARAP} applied
different $\ell_p$ sparsity norms to the as-rigid-as-possible
energy~\cite{sorkine2007ARAP} to create various deformation styles.
Recently, Chen et al.~\shortcite{chen2017L1ARAP} used $\ell_{2,1}$ regularization on vertex
positions to locally control the deformation.  However, their
direct use of the $\ell_{2,1}$ norm will create artifacts when the control point is
not on the boundary of the shape (see \reffig{octocat_bird_vs_all}(b) and \reffig{crocodile_bird_vs_all}(b)).  Moreover,
their method requires one ADMM solve in each global iteration, which 
renders the optimization less efficient and slow in runtime.
Also, their framework is
limited to 2D deformation with ARAP energy only, while our framework
generalizes across dimensions and a variety of energy models.

Our algorithm, inspired by sparsity-seeking regularizers such as
that used by \changed{\cite{chen2017L1ARAP, fan2001scad}}, addresses these shortcomings.  
In particular, we propose a simple and novel sparsity-inducing norm
that eliminates artifacts arising from the $\ell_{2,1}$ norm, and our efficient
optimization scheme leads to interactive performance.
 
\vspace{-2mm}
\section{Overview}

The main idea of our method is to use a novel regularization term to produce local deformation with a dynamic region of influence (ROI).
Our method takes a triangle/tetrahedral mesh (or a 1D polyline) and a set of selected vertices as control handles as input.
The output of our method is a deformed shape where the deformation is both \emph{local} and \emph{natural} and the ROI is automatically adaptive to the deformation.
Here the ``locality'' implies that a handle only dominates its nearby areas without affecting the regions far away.

\vspace{-2mm}
\subsection{\Lossnameshort{} Regularization}
\label{sec:bird-loss-overview}

We suggest a sparsity-inducing regularization term to produce natural local deformation.
This regularizer is applied per-vertex to $\mV_i - \tlmV_i$ to bias each vertex deformed position $\mV_i$ to exactly match its initial rest position $\tlmV_i$ except in isolated regions of the shape.
The most obvious choice for this regularization would be to enforce sparsity either with an $\ell_1$-norm, or with a \textit{group lasso} / \textit{\changed{$\ell_{2,1}$} regularization} defined as $\sum_i \Vert \mV_i \Vert_2$ as in \cite{chen2017L1ARAP}.

However, direct use of a \changed{$\ell_{2,1}$} regularization term leads to artifacts, due to the fact that the \changed{$\ell_{2,1}$}-norm competes with the elastic energy by dragging \emph{all} the vertices towards their original positions.
This results in undesired distortion near the deformation handles (see \reffig{octocat_bird_vs_all}(b) and \reffig{crocodile_bird_vs_all}(b)), making the deformed region look unnatural.
Previous \changed{$\ell_{2,1}$}-based methods \cite{chen2017L1ARAP} focus specifically on ARAP-like deformation, and attempt to alleviate this artifact by adding a Laplacian smoothness term and a weighting term based on biharmonic distance.
Unfortunately, as seen in \reffig{octocat_bird_vs_all} at the ends of the octocat's tentacles, artifacts can arise in regions quite close to the deformation handles, and there is not necessarily any setting of these parameters which alleviates these artifacts without oversmoothing the entire deformation.

\changed{Inspired by ``folded concave'' losses in statistical regression \cite{fan2001scad,10.1214/09-AOS729} and the use of the $\ell_{2,1}$ loss in deformation}
\changed{\cite{chen2017L1ARAP}, we propose using a smoothly clipped group $\ell_1$-norm as our locality-inducing regularization. We call it a \lossname{} for ``smoothly clamped $\ell_1$ loss'' (see the inset).}

\changed{We adopt a simple implementation for our \lossname{}}:
\begin{align} 
\Vert \vx \Vert_{\mlossname}
=
\left\{\begin{matrix} 
	\Vert \vx \Vert_2 - \frac{1}{2s} \Vert \vx \Vert_2^2 & \Vert \vx \Vert_2 < s \\
	\frac{1}{2} s & \Vert \vx \Vert_2 \geq s
\end{matrix}\right. ,
\label{equ:bird_loss_function}
\end{align}
\begin{wrapfigure}[6]{r}{1.33in}
	\raggedleft
  \vspace{-5pt}
	\includegraphics[width=1.31in, trim={6mm 0mm 1mm 0mm}]{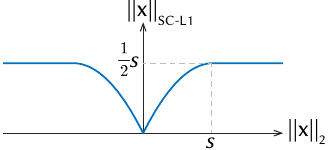} 
	\label{fig:bird_loss_func}
\end{wrapfigure} 
where $s$ is the threshold distance beyond which the regularizer is disabled (this is equivalent to a group-variant of the MCP loss in \cite{10.1214/09-AOS729}, but we use the term ``\lossnameshort'' to emphasize that the minimax concave property is not critical for localized deformation).
This function is continuously differentiable and piecewise smooth, and admits a proximal shrinkage operator free of local minima.
Near the origin the \lossname{} function acts like the group $\ell_1$-norm, which drives $\Vert \vx \Vert_{\mlossname}$ towards $0$ \changed{in a sparse-deformation-seeking manner}.
When $\Vert \vx \Vert_2 \geq s$, the \lossname{} function value is a constant and has no penalty on $\Vert \vx \Vert_2$.
\changed{For a detailed comparison between our \lossname{} function and other alternatives, please see Sec.5 of the supplementary material. }

\subsection{Local Deformation Energy}
\label{sec:energy-overview}

We denote $\mV$ as a $\vert \mV \vert \times d$ matrix of vertex positions at the deformed state, and $\tlmV$ as a $\vert \mV \vert \times d$ matrix containing rest state vertex positions.

The total energy for our local deformation is as follows:
\begin{subequations}
\begin{align}
  \underset{\mV}{\minimize} \quad &
  \underbrace{E(\mV)}_{\text{Elasticity}}  +  
  \sum_{i \in V} \underbrace{\changed{w}  a_i \| \mV_i - \tlmV_i\|_{\mlossname}}_{\text{Locality}}, \\
  \st \quad & \mV_s = \vp_s,     \quad \text{(position constraint)} & \\
            & \mA_k \mV_t = \vb_k,  \forall t \in \MS_k              \quad \text{(affine constraint)} &
\end{align} \label{equ:general_form}
\end{subequations} 
The first term is an elasticity energy of choice, and can be selected independent of the locality regularization.
The second term is the novel ``\lossname{}'' term on the vertex position changes, which measures the locality of the deformation. 
$a_i$ is the barycentric vertex area of the $i$-th vertex, which ensures the consistency of the result across different mesh resolutions for the same \changed{constant} $\changed{w}$.
To enable more user control, position constraints and optional affine constraints can be added on selected vertices to achieve different deformation effects.
$s$ denotes the indices of the vertices with the position constraint, and we call these vertices ``handles''.
$\MS_k$ is the $k$-th set of vertex indices where an affine constraint is added.
For simplicity, we omit the position constraints and affine constraints in the discussion below, as they can be easily intergrated to the system by removing the corresponding degrees of freedom and using Lagrange multiplier method (see Eq(1) in \cite{wang2015bhc}).

Inspired by the local-global strategy in \cite{brown2021wrapd}, our local deformation energy (\refequ{general_form}) can be rewritten as:
\begin{subequations}
\begin{align}
  \underset{\mV, \{\mX_j\}}{\minimize} \quad &
  \sum_{j} \underbrace{E(\mX_j)}_{\text{Elasticity}}  +  
  \sum_{i \in V} \underbrace{\changed{w}   a_i \| \mV_i - \tlmV_i\|_{\mlossname}}_{\text{Localness}}, \\
  \st \quad & \mX_j = \sym{\mD_j \mV}, \forall j,
\end{align} \label{equ:local_global_form}
\end{subequations}
where $\mD_j$ is the selection matrix for edges of the $j$-th vertex or element.
$\sym{\mF}$ denotes the symmetric factor $\mS$ computed using the polar decomposition $\mF = \mR \mS$, where $\mF$ is the deformation gradient.
Thus $\mX_j$ is the symmetric factor of deformation gradient of the $j$-th vertex or element.
(Note that in general $\sym{\mF} \ne \frac{1}{2}(\mF + \mF^\top)$.) 
The goal of using $\sym{}$ here is to ensure the local coordinates $\mX_j$ are invariant to rotations as well as translations.
For details, please see \cite{brown2021wrapd}.

\section{Optimizing With ADMM}
\label{sec:admm-overview}

\begin{algorithm}[t]
  \caption{Three-block ADMM Overview}\label{alg:admm-overview}
  \KwIn{A triangle or tetrahedral mesh $\tl{\mV},\mT$} 
  \KwOut{Deformed vertex positions $\mV$} 
  $\mV \gets \tl{\mV}$ \\
  \While{not converged} {
    $\mX_i \gets$ \textit{local\_step\_X}$(\mV, \tl{\mV})$ \Comment{local step 1} \\
    $\mZ_i \gets$ \textit{local\_step\_Z}$(\mV, \tl{\mV})$ \Comment{local step 2} \\
    $\mV \gets$ \textit{global\_step}$(\mX_i, \mZ_i, \tl{\mV})$ \Comment{global step} \\
    \gray{$\mU_i \gets$ \textit{dual\_update}$(\mZ_i, \mV_i, \tl{\mV}_i)$ \Comment{dual update 1}} \\
    \gray{$\mW_i \gets$ \textit{dual\_update}$(\mX_i, \mV_i, \tl{\mV}_i)$ \Comment{dual update 2}}
  }
\end{algorithm}

A natural way to minimize this energy is to use the alternating direction method of multipliers \cite{boyd2011ADMM} for the sparsity term, and to use a local-global update strategy for the elasticity term.
However, previous \changed{$\ell_{2,1}$}-based methods \cite{chen2017L1ARAP} apply these two strategies separately in their local and global steps, resulting in an inefficient optimization scheme.
As they only support 2D ARAP energy, we discuss further in \refsec{arap-bird-loss}.

We propose a new way to efficiently minimize such energies in \refequ{local_global_form} by combining the sparsity-targeted ADMM with the elasticity-focused local-global strategy.
We minimize our energy (\refequ{local_global_form}) using a three-block alternating direction method of multipliers scheme \cite{boyd2011ADMM} following the local-global update strategy. 
Our first local step, finding the optimal symmetric factor $\mX_i$ of the deformation gradient, can be formulated as a minimization problem on \changed{its singular values}. 
Our second local step, minimizing the \lossname{} term for each $\mZ_i$, can be solved using a shrinkage step. 
Our global step, updating vertex positions $\mV$, is achieved by solving a linear system. 
We provide an overview of our three-block ADMM scheme in \refalg{admm-overview}.

\subsection{Example I: Local ARAP Energy}
\label{sec:arap-bird-loss}

We begin by considering how to minimize an as-rigid-as-possible (ARAP) energy \cite{sorkine2007ARAP} when combined with a \lossname{} regularizer.
With an ARAP elastic energy, the total energy (\refequ{local_global_form}) for our local deformation is as follows:
\begin{align}
  \underset{\mV,\{\mR_i\}}{\minimize} \quad \sum_{i \in V} \quad 
  \underbrace{\frac{1}{2} \| \mR_i \mD_i - \tl{\mD}_i \|_{\mW_i}^2}_{\text{ARAP}}  +  
  \underbrace{\changed{w} a_i \| \mV_i - \tlmV_i\|_{\mlossname}}_{\text{Localness}}, \label{equ:bird_arap_energy}
\end{align}
where $\mR_i$ is a $d \times d$ rotation matrix, $W_i$ is a $\vert \MN(i) \vert \times \vert \MN(i) \vert$ diagonal matrix of cotangent weights, $\tl{\mD}_i$ and $\mD_i$ are $3 \times \vert \MN(i) \vert$ matrices of "spokes and rims" edge vectors of the $i$-th vertex at the rest and deformed states respectively.  $\Vert \mX \Vert_{\mW_i}^2$ denotes $\Tr(\mX^\top \mW_i \mX )$. 
Here we use $\mR_i$ to denote $\mX_i$, since we drive the deformation gradient towards a rotation matrix in ARAP energy.

Previous method \cite{chen2017L1ARAP} optimizes the \changed{$\ell_{2,1}$} version of \refequ{bird_arap_energy} in a less efficient way.
Their local step optimizes over per-vertex rotation $\mR_i$ and their global step minimizes over vertex 
\begin{wrapfigure}[6]{r}{1.33in}
	\raggedleft
  \vspace{-1pt}
	\includegraphics[width=1.31in, trim={6.5mm 0mm 0mm 3mm}]{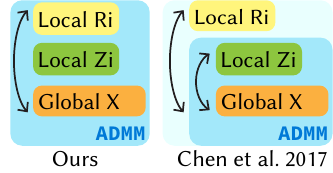} 
	\label{fig:admm_3_block}
\end{wrapfigure} 
positions $\mV$ using a two-block ADMM scheme.
This leads to an expensive optimization with a full ADMM optimization in \emph{each} global step, making their method too slow for interactive usage. 
In contrast, applying our new three-block ADMM scheme to the local ARAP energy results in a much more efficient solver, which is one ADMM optimization itself (see the inset, where the blue regions denote an ADMM optimization).
We further show the pseudocode of our three-block ADMM for local ARAP energy in Suppl. Alg.1.

More concretely, by setting $\mZ_i = \mV_i - \tlmV_i$, we can further rewrite \refequ{bird_arap_energy} as
\begin{subequations}
\begin{align}
  \underset{\mV,\{\mR_i\},\mZ}{\minimize} & \quad \sum_{i \in V} \quad \frac{1}{2} \| \mR_i \mD_i - \tl{\mD}_i \|_{\mW_i}^2  +  \changed{w} a_i \| \mZ_i\|_{\mlossname}, \\
  \st & \quad \mZ_i = \mV_i - \tl{\mV}_i, \quad \forall i.
\end{align} \label{equ:bird_arap_energy_admm}
\end{subequations}
The above minimization problem can be solved efficiently using the following ADMM update steps:
\begin{subequations}
\begin{alignat}{1}
  & \mR_i^{k+1} \leftarrow \argmin_{\mR_i \in \text{SO(3)}} \;   \frac{1}{2} \| \mR_i \mD_i - \tl{\mD}_i \|_{\mW_i}^2  \label{equ:admm_local_argminX}\\
  & \mZ_i^{k+1} \leftarrow \argmin_{\mZ_i} \; \changed{w} a_i \| \mZ_i \|_{\mlossname} + \frac{\rho}{2} \| \mV_i^{k+1} - \tl{\mV}_i - \mZ_i + \mU_i^k  \|_2^2  \label{equ:admm_local_argminZ}\\
  & \mV^{k+1} \leftarrow \argmin_{\mV} \;  \mW (\mV^{\top}\mL\mV - \mB^{\top}\mV)  + \frac{\rho}{2} \| \mV - \tl{\mV} - \mZ^k + \mU^k \|_2^2  \label{equ:admm_global_argminV}\\
  & \mU_i^{k+1} \leftarrow \mU_i^k + \mV^{k+1}_i - \tl{\mV}_i - \mZ_i^{k+1} \label{equ:admm_global_updateU}
\end{alignat} \label{equ:arap_admm}
\end{subequations}
Here $\rho$ is a fixed penalty parameter.
For a detailed derivation of the ADMM update, please see Sec. 2 of the supplementary material.

The various steps in this ADMM-based algorithm are computed as follows:

For \textbf{updating $\mR_i$}, local step 1 (\refequ{admm_local_argminX}) is an instance of the Orthogonal Procrustes problem, which can be solved in the same way as the rotation fitting step in \cite{sorkine2007ARAP}.
The optimal $\mR_i$ can be computed as $\mR_i^{k+1} \leftarrow \mathcal{V}_i \mathcal{U}_i^{\top}$ from the singular value decomposition of $\mM_i=\mathcal{U}_i \Sigma_i \mathcal{V}_i^{\top}$, where $\mM_i= \mD_i \tl{\mD}_i^\top$.

For \textbf{updating $\mZ_i$}, following the derivation of the proximal operator of \lossname{} in Sec. 1 of the supplementary material, our local step 2 (\refequ{admm_local_argminZ}) is solved using a \lossname{}-specific shrinkage step:
\begin{align}
  &\mZ_i^{k+1} \leftarrow S_{\changed{w} a_{i}}^{k}\left(\mV_i - \tl{\mV_i}  + \mU_i \right) \\
  &\mathcal{S}_{\changed{w} a_i}(\mathrm{\vx})= \begin{cases} \left(\frac{\rho s - \changed{w} a_i s / \Vert \vx \Vert_2}{\rho s - \changed{w} a_i}\right)_{+} \vx,  \text{ if } \Vert \vx \Vert_2 \le s \\
  \vx, \space \text{ otherwise}
  \end{cases}
\end{align}
To avoid local minima in the shrinkage step, this assumes $\rho$ is set to satisfy $\rho > \frac{\text{max}(\changed{w} a_i)}{s}$ (see Sec. 1 of the supplementary material).

For \textbf{updating $\mV$}, the global step (\refequ{admm_global_argminV}) can be achieved by solving a linear system:
\begin{align}
  (\mL + \rho \mI)\mV = \mB + \rho (\tl{\mV} + \mZ^k - \mU^k),
\end{align}
where the Laplacian $\mL$ and $\mB$ are defined in the same way as the global step (Eq. 9) in \cite{sorkine2007ARAP}.
For fixed $\rho$ an efficient implementation is obtained by precomputing and storing the Cholesky factorization of $\mL + \rho \mI$.

\subsection{Example II: Local Neo-Hookean Energy} \label{sec:neo-hookean-bird-loss}

Our local deformation scheme can be further extended to physics-based elasticity energies, e.g., Neo-Hookean energy.
Using the Neo-Hookean energy as our elasticity energy and following the framework of \cite{brown2021wrapd}, the optimization problem in \refequ{local_global_form} can be written as follows:
\begin{subequations}
\begin{align}
  \underset{\mV, \{\mX_j\}, \{\mZ_i\}}{\minimize} \quad &
  \sum_{j \in T} \underbrace{E_{\text{nh}}(\mX_j)}_{\text{Neo-Hookean}}  +  
  \sum_{i \in V} \underbrace{\changed{w}   a_i \| \mV_i - \tlmV_i\|_{\mlossname}}_{\text{Locality}}, \\
  \st \quad & \mX_j = \sym{\mD_j \mV}, \forall j,
\end{align} \label{equ:neo_hookean_bird_energy}
\end{subequations} 
where $T$ denotes all the elements.

Similarly, by introducing $\mZ_i = \mV_i - \tlmV_i$, we can minimize our local Neo-Hookean energy using ADMM. 
For a detailed derivation, please see Sec. 3 of the supplementary material.

The ADMM update (\refalg{admm-overview}) for the above minimization problem is as follows:
\begin{subequations}
\begin{alignat}{1}
  &  \mX_j^{k+1} \leftarrow \argmin_{\mX_j} \;  E_{\text{nh}}(\mX_j) + \frac{\gamma}{2} \| \sym{\mD_j \mV} - \mX_j + \mW_j \|_2^2   \label{equ:NH_admm_local_argminX}\\
  &  \mZ_i^{k+1} \leftarrow \argmin_{\mZ_i} \; \changed{w} a_i \| \mZ_i \|_{\mlossname} + \frac{\rho}{2} \| \mV_i^{k+1} - \tl{\mV}_i - \mZ_i + \mU_i^k  \|_2^2  \label{equ:NH_admm_local_argminZ}\\
  &  \begin{matrix} \mV^{k+1} \leftarrow \underset{\mV}{\argmin} \;  \sum_{j \in T} \frac{\gamma}{2} \| \sym{\mD_j \mV} - \mX_j + \mW_j \|_2^2 \\ \qquad \qquad \qquad  + \sum_{i \in V} \frac{\rho}{2} \| \mV - \tl{\mV} - \mZ^k + \mU^k \|_2^2 \end{matrix}  \label{equ:NH_admm_global_argminV}\\
  &  \mW_j^{k+1} \leftarrow \mW_j^k + \sym{\mD_j \mV} - \mX_j \label{equ:NH_admm_global_updateW} \\
  &  \mU_i^{k+1} \leftarrow \mU_i^k + \mV^{k+1} - \tl{\mV}_i - \mZ_i^{k+1} \label{equ:NH_admm_global_updateU}
\end{alignat} \label{equ:neo_hookean_admm}
\end{subequations}
Here $\rho$ and $\gamma$ are fixed penalty parameters.

The local step 2 (updating $\mZ_i$) and the global step (updating $\mV$) can be solved in the same way as the local ARAP energy (see \refsec{arap-bird-loss}).

For \textbf{updating $\mX_i$}, local step 1 (\refequ{NH_admm_local_argminX}) can be solved by performing the energy minimization on the singular values of $\sym{\mD_j \mV} + \mW_j$.
This is a proximal operator of $E_{\text{nh}}$ at the rotation-invariant $\sym{\mD_j \mV} + \mW_j$. 

Let us denote the proximal operator of $E_{\text{nh}}$ and the singular value decomposition of $\sym{\mD_j \mV} + \mW_j$ as:
\begin{align}
 & \operatorname{prox}_{E_{\text{nh}}}(\mX_j) = E_{\text{nh}}(\mX_j) + \frac{\gamma}{2} \| \sym{\mD_j \mV} + \mW_j - \mX_j \|_2^2  \\
 & \sym{\mD_j \mV} + \mW_j = \mathcal{U}_j \Sigma_j \mathcal{V}_j^{\top}
\end{align}
where $\gamma$ is the augmented Lagrangian parameter for $\mX_j$.

As shown by \cite{brown2021wrapd}, we can compute the optimal $\mX_j$ as:
\begin{align}
 & \Sigma_j^{k+1} \leftarrow \argmin_{\Sigma_j}  \;  \operatorname{prox}_{E_{\text{nh}}}(\mathcal{U}_j \Sigma_j \mathcal{V}_j^{\top})  \\
 & \mX_j \leftarrow \mathcal{U}_j \Sigma_j^{k+1} \mathcal{V}_j^{\top}
\end{align}

Specifically, one can compute the SVD of $\sym{\mD_j \mV} + \mW_j$ and perform the minimization of $\operatorname{prox}_{E_{\text{nh}}}$ only on its singular values, while keeping singular vectors unchanged.
The above optimization of singular values $\Sigma_j^{k+1}$ can be performed using an L-BFGS solver.

\begin{algorithm}[t]
  \caption{Three-block ADMM for local ARAP Energy}\label{alg:arap-admm}
  \KwIn{A triangle or tetrahedral mesh $\tl{\mV},\mT$} 
  \KwOut{Deformed vertex positions $\mV$} 
  $\mV \gets \tl{\mV}$ \\
  \While{not converged} {
    $\mR_i \gets$ \textit{local\_step\_X}$(\mV, \tl{\mV})$ \Comment{local step 1} \\
    $\mZ_i \gets$ \textit{local\_step\_Z}$(\mV, \tl{\mV})$ \Comment{local step 2} \\
    $\mV \gets$ \textit{global\_step}$(\mX_i, \mZ_i, \tl{\mV})$ \Comment{global step} \\
    \gray{$\mU_i \gets$ \textit{dual\_update}$(\mZ_i, \mV_i, \tl{\mV}_i)$ \Comment{dual update 1}}
  }
\end{algorithm}

\subsection{Extension to Other Elastic Energies} \label{sec:bird-loss-extension}

Our algorithm can easily generalize across different dimensions and material models.
Switching the material model only requires a change on the minimization problem in the local step 1 $\argmin_X$, which can be optimized over the singular values of the symmetric factor $\mX_i$ of the deformation gradient.

\setcounter{figure}{3}
\begin{figure}[t]%
  \centering
  \includegraphics[width=0.85\linewidth]{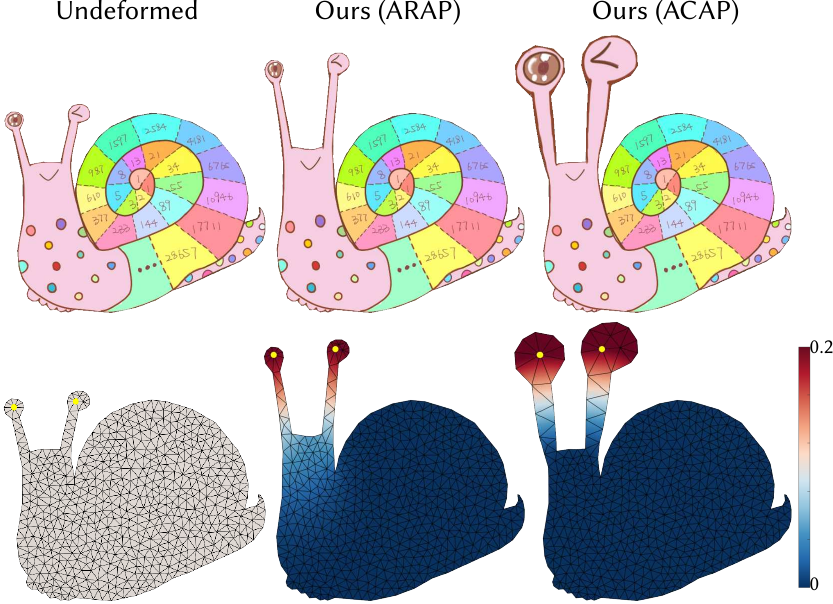}
  \caption{Our method automatically choose a natural ROI based on the elastic energy in use.
  Here we use the same parameter settings for both the local ARAP and local ACAP energies. 
  With the local ACAP energy, we have a smaller ROI than the case of local ARAP energy 
  as the ACAP energy allows for local scaling.}
  \label{fig:fibonacci_snail_acap}
\end{figure}

\setcounter{figure}{4}
\begin{figure}[t]%
  \centering
  \includegraphics[width=0.9\linewidth]{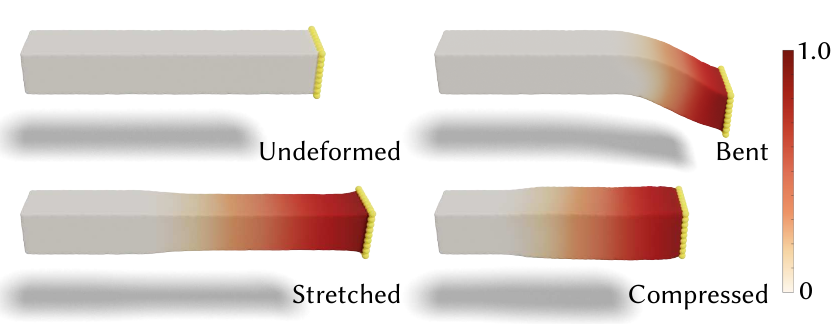}
  \caption{\changed{
    Given the \emph{same} handle offset magnitude, a "natural" ROI size also depends on the way the handles are moved, which is 
    more complex than simply growing the ROI proportional to the handle displacement. 
    Here we moved the handles (yellow) with the \emph{same} offset magnitude 1.0 towards the bottom, left and right respectively, resulting different ROIs for the 
    \emph{same} handle offset magnitude.
  }}
  \label{fig:bar_stretch_compress_bend}
\end{figure}

\subsubsection{As-Conformal-As-Possible Energy} 

For editing tasks where users intend to locally scale the geometry while preserving the texture, it's desirable to constrain the angle preservation, or conformality (see \reffig{fibonacci_snail_acap} and \reffig{bird_acap}).
We can adapt the ARAP energy to as-conformal-as-possible (ACAP) energy \cite{bouaziz2012shapeUp} by allowing local scaling:
\begin{align}
  E_{\text{ACAP}}(\mV) = \sum_{k \in T} \sum_{i,j \in \MN(k)} \frac{w_{ij}}{2} \Vert s_k \mR_k \tl{\vd}_{ij} - \vd_{ij} \Vert_2^2
\end{align}
where $s_k$ is a scalar controlling the scaling of the local patch and can be computed analytically (\changed{see Sec.4 of the supplementary material}).

\subsubsection{Cloth}

Our method also generalizes to higher co-dimensional settings, such as deformable thin sheets and cloth in $\mathbb{R}^3$.
We model the cloth deformation using ARAP elasticity (\refequ{bird_arap_energy}), hard strain limiting, and quadratic bending resistance \cite{bergou2006bending}. 

\subsubsection{1D Polyline}

Our algorithm can be also extended to the local editing of 1D polyline in vector graphics.
The deformation of a polyline can be modeled using the ARAP energy (\refequ{bird_arap_energy}) with uniform weights.

\section{Results}
\label{sec:result}

We evaluate our method by 
comparing it against existing local deformation tools
and showcasing its extension to various elastic energies.
All the colormaps in our figures visualize the vertex displacement with respect to the rest shape.
The accompanying video also includes several animation examples generated using our local deformation tool.

We implement a 2D version of our method in MATLAB with gptoolbox
\cite{gptoolbox}, and a 3D version in C++ with libigl \cite{libigl} based on
the WRAPD framework \cite{brown2021wrapd}.  We also implement the 2D version
of our method in C++ for runtime evaluation and comparison. Benchmarks are
performed using a MacBook Pro with an Apple M2 processor and 24GB of RAM for 3D
and a Windows desktop with an i9-9900K 3.60 GHz CPU for 2D.  Table 2 in the supplementary material shows the performance statistics and relevant parameters of all our
examples.

\paragraph{Quality}
We compare our methods against other local editing tools, including \textbf{i)} $\ell_1$-based
deformation~\cite{chen2017L1ARAP}, \textbf{ii)} regularized Kelvinlets~\cite{DeGoes2017Kelvinlets} and 
\textbf{iii)} 
biharmonic coordinates~\cite{wang2015bhc}.
Among them, the use of \changed{$\ell_{2,1}$}-norm regularization~\cite{chen2017L1ARAP}
causes artifacts (see \reffig{octocat_bird_vs_all}-b and \reffig{crocodile_bird_vs_all}-b).
Regularized Kelvinlets technique \cite{DeGoes2017Kelvinlets} deforms a shape
based on Euclidean distances, thus not shape-aware, creating artifacts when two 
disjoint parts are close in Euclidean space but far away geodesically  
(see the blue region in \reffig{octocat_bird_vs_all}-c and the teeth area in
\reffig{crocodile_bird_vs_all}-c).  
Methods based on biharmonic coordinates, such as \cite{wang2015bhc}, usually require careful
placement of additional fixed control points to pre-determine the ROI.  
The latter two methods do not minimize any elastic energy in the deformation process, 
and thus their deformations are more susceptible to shape distortion (see \reffig{crocodile_bird_vs_all}).
We additionally compare our method against the \emph{sparse} deformation method~\cite{lin2012LpARAP},
which directly introduces a sparsity-induced norm in ARAP energy.  
As shown in \reffig{bar_2d_Lp}, the resulting deformation is
\emph{sparse} but not \emph{local}, thus requiring the setup of 
additional fixed constraints.

In contrast, our method produces the deformation which is local, natural, and
shape-aware; it automatically adapts the ROI without the need for careful control
primitive setup.

\paragraph{Efficacy}
We illustrate the ROI adaptation of our method in different situations: 
It adapts to different energy models\textemdash for example, the local ACAP has a smaller 
ROI than the local ARAP energy as the \changed{former} allows for local scaling (see \reffig{fibonacci_snail_acap}).
It also adapts to different extents of deformation.
As \reffig{bar_3d_roi} demonstrates, the ROI gradually increases as the deformation of the bar
becomes larger.

One can configure the local deformation style by choosing
various elastic energy models. For example, the local
Neo-Hookean energy leads to deformation that preserves volume, while 
the local ARAP energy is volume agnostic \changed{(see \reffig{pig_arap_vs_nh})}.
The deformation can be further tuned by introducing additional affine constraints\textemdash
for instance, to enable the character to wave hands (\reffig{man_2d_affine})
or the crocodile to open its mouth (\reffig{crocodile_bird_vs_all})\textemdash
in a natural way.

\paragraph{Performance}
In terms of performance, our solver is able to efficiently
minimize the energy at interactive rate, while the method of
\cite{chen2017L1ARAP} is too slow to run in realtime.  Because
their method only supports 2D ARAP energy, in Table
1 of the supplementary material, we evaluate the runtime of our method 
(using both the \lossname{} and \changed{$\ell_{2,1}$} loss) and 
\cite{chen2017L1ARAP} on a 2D ARAP local
energy and across different mesh resolutions and deformations.  
Measured with the same convergence threshold, our method runs orders of magnitude faster than
\cite{chen2017L1ARAP}, achieving roughly $1000\times$ speedup for small deformation and
$100\times$ speedup for large deformation.

\paragraph{Extensibility}
Our method can easily generalize to other dimensions and material models,
such as the ACAP deformation, cloth deformation, and 1D polyline deformation
(see \reffig{bird_acap} and \reffig{fibonacci_snail_acap}). The local ACAP
energy enables local scaling and better preserves the texture
around the deformed region.  In \reffig{polyline_1d}, the
user can interactively edit a polyline and naturally recovers its rest
shape, which is a desirable feature by the users.
In \reffig{cloth_posing}, to deform a cloth in a physically plausible way,
the deformation locality is particularly useful, as otherwise a local edit of the cloth may cause 
a global change leading to unexpected intersections with other objects. 
Lastly, to demonstrate our method in a more complex scenario, an editing session involving multiple objects
and clothes is shown in \reffig{teaser}.

\vspace{-2pt}
\section{CONCLUSION \& FUTURE WORK}

We describe a regularization based on an ``\lossname{}'' which provides an effective and simple to implement tool for localizing an elastic energy driven deformation to only those regions of a shape being manipulated by a user.
The region of influence induced by our method naturally adapts to the geometry of the shape, the size of the deformation, and the elastic energy being used.
Furthermore \lossnameshort{} regularization is generic enough to be applied to a wide range of shapes and elastic energies, including 1D, 2D, 3D and cloth finite element deformation, and is fast enough to be used in real-time.
Our proposed approach offers several benefits for shape manipulation: It avoids undesired movement in far-off regions of a shape when only one part is being moved by the user, it allows parts of a shape to be deformed with direct manipulation without a pre-rigging step, and avoids the visual artifacts of previous work.

There remain several issues related to localized shape deformation not addressed by our method.
Firstly our regularization is applied independently per-vertex, which makes it difficult to apply to splines, NURBS, or even meshes with highly irregular element sizes, which we mark an important direction for future work.
In addition, since we use an ADMM method in the optimization, our approach suffers from the common shortcomings of applying ADMM to non-convex energies, including lack of convergence guarantees and slow convergence when high precision is required.
Exploration of other optimization algorithms alleviating these issues is another useful future direction.
\changed{
Finally, although it is out of scope for our work here, we note in particular the usefulness of incorporating localized elastic energy deformation into sculpting workflows for artists.
This involves a number of facets: Choosing the correct elastic energy to achieve an artistic effect, providing an intuitive UI to adjust the scale of the ROI (for instance by adjusting $w$ and $s$, see \reffig{crocodile_param_s_w} and the supplementary video), and in ensuring that our tool integrates well with other sculpting tools.
This is particularly useful when handling large ``freeform'' deformations, as the elastic energy will tend to fight against such deformations, making other tools more suitable.
One simple idea for this is to simply reset the rest shape after each click-and-drag, since each deformation step is then independent of the others, and one could switch between our method and others at each step.
We have found this mode of interaction to be useful even when only using our method, as it leads to a simple sculpting-style interface, and we include some examples in the supplementary video.
}

\begin{acks}
This work is funded in part by National Science Foundation (1910839).
 We especially thank George E. Brown for sharing the WRAPD implementation and the help with setting up experiments.
 We thank Jiayi Eris Zhang and Danny Kaufman for sharing the undeformed scene geometry;
 Lillie Kittredge and Huy Ha for proofreading; 
 Rundi Wu for the help with rendering; 
 all the artists for sharing the 2D and 3D models and 
 anonymous reviewers for their helpful comments and suggestions.
\end{acks}

\bibliographystyle{ACM-Reference-Format}
\bibliography{sections/reference.bib}

\newpage

\setcounter{figure}{5}
\begin{figure*}[t]%
  \centering
  \includegraphics[width=\textwidth]{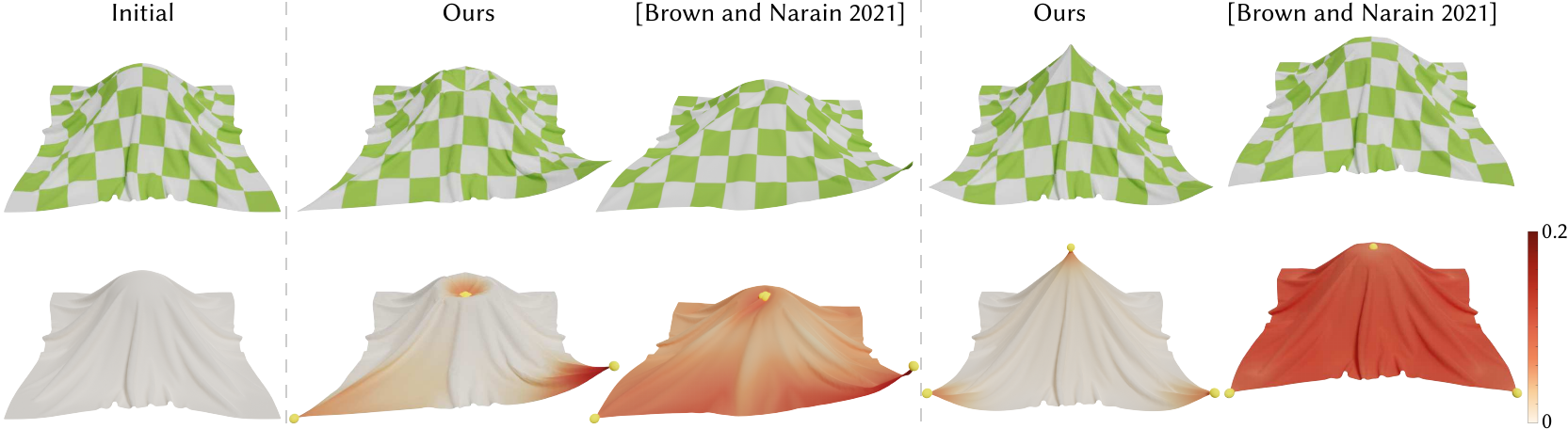}
  \vspace{-2mm}
  \caption{Our method enables the user to locally edit a simulated cloth in an interactive fashion, without the need of rerunning the simulation. 
    In contrast, quasi-static deformations generated by directly moving the control points \cite{brown2021wrapd} have global effects,
easily deviating the edit from the initial shape in a distinct way.  
\textbf{Bottom row:} To show the extent to which the handles affect cloth vertex positions, 
we colormap the vertex displayment from the input cloth.
The handles moved by the user are in yellow.
(Undeformed geometry thanks to \cite{zhang2022mgCloth})
}
  \label{fig:cloth_posing}
\end{figure*}

\setcounter{figure}{6}
\begin{figure*}[h]%
  \centering
  \includegraphics[width=\textwidth]{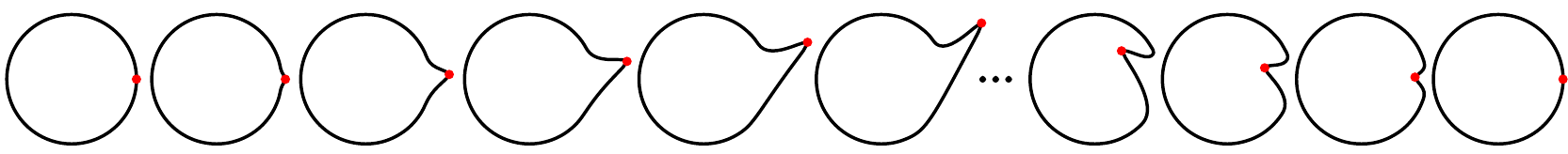}
  \caption{Our method also generalizes to 1D polyline editing. Our method enables the deformed shape to naturally return to the rest shape when they are close enough. }
  \label{fig:polyline_1d}
\end{figure*}

\setcounter{figure}{7}
\begin{figure*}[h]%
  \centering
  \includegraphics[width=\textwidth]{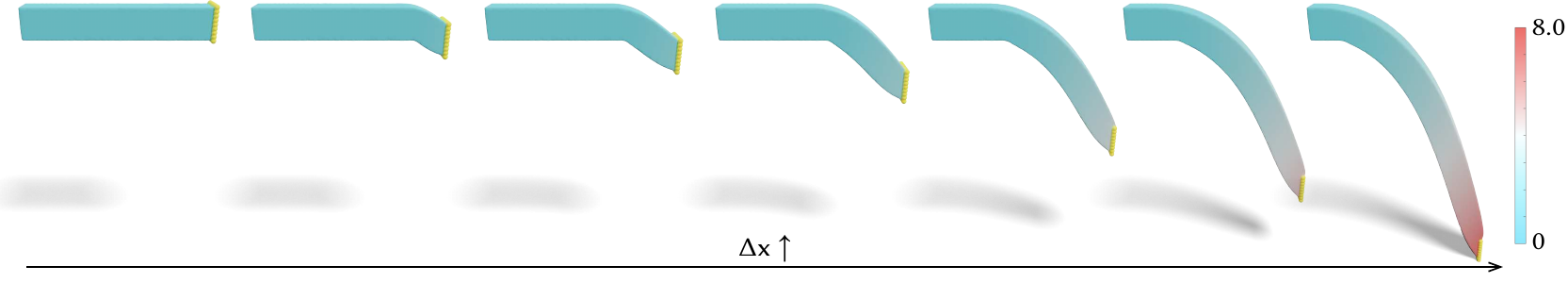}
  \caption{\changed{Our algorithm naturally enables an adaptive region of influence
  under different deformations.  Only the vertices (yellow) on the rightmost
  end are selected as handles. From left to right, the offset $\Delta x$ of the
  rightmost end is 0.0, 0.5, 1.0, 2.0, 4.0, 6.0 and 8.0 respectively.} }
  \label{fig:bar_3d_roi}
\end{figure*}

\setcounter{figure}{8}
\begin{figure*}[h]%
  \centering
  \includegraphics[width=\textwidth]{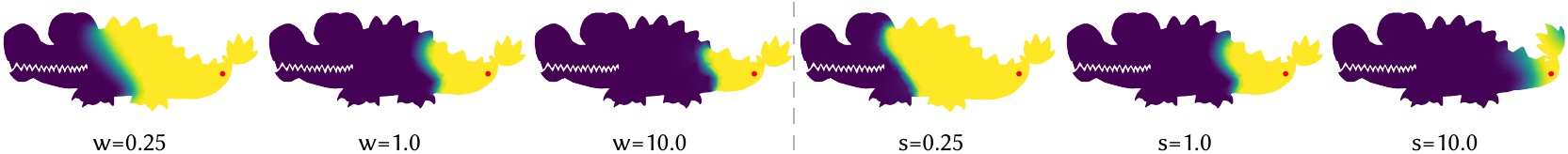}
  \caption{\changed{We show how the ROI (yellow) changes when adjusting $s$ or $w$, while keeping the rest the same. Here the handles are highlighted in red.
  In general, we use $w$ to control the scale of the ROI, and set $s$ to a small factor of the size of the shape and then leave it alone. } }
  \label{fig:crocodile_param_s_w}
\end{figure*}

\setcounter{figure}{9}
\begin{figure}[t]%
  \centering
  \includegraphics[width=\linewidth]{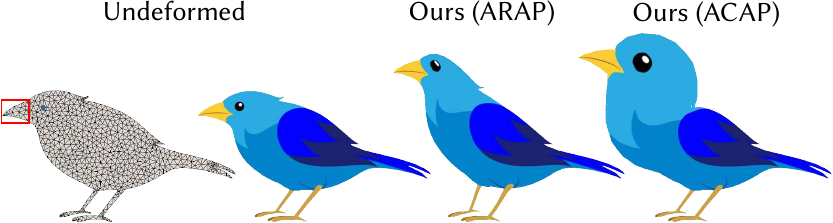}
  \caption{
      As-conformal-as-possible (ACAP) energy encourages
      conformality, thus better preserving the color texture by
      allowing local scaling, while the as-rigid-as-possible (ARAP) energy favors
      preserving the rigidity.  Here the handle (in blue) is
      placed around the eye and an affine constraint is added
      to the red region. 
  }
  \label{fig:bird_acap}
  \vspace{0pt}
\end{figure}

\setcounter{figure}{10}
\begin{figure}[h]%
  \centering
  \includegraphics[width=\linewidth]{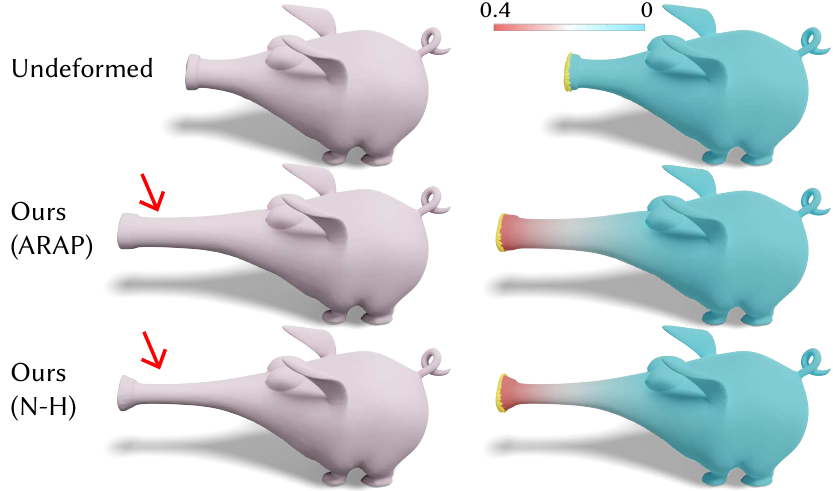}
  \caption{\changed{ARAP v.s. Neo-Hookean energy with \lossname{}. Only the vertices (yellow) on the nose of the pig are selected as handles. Note that Neo-Hookean version exhibits the volume-preservation property. }}
  \label{fig:pig_arap_vs_nh}
  \vspace{0pt}
\end{figure}

\setcounter{figure}{11}
\begin{figure}[h]%
  \centering
  \includegraphics[width=\linewidth]{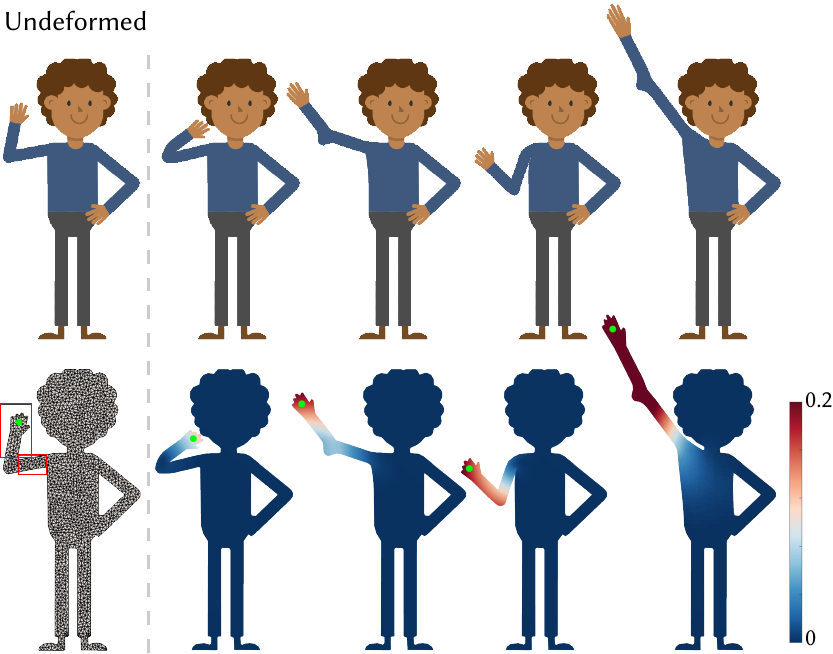}
  \caption{Our method supports adding both positional constraints (green) and affine constraints (red) on vertices. The region of influcence in the deformed shape naturally adapts to the resulting deformation. 
  \newline
  \tiny{Black Man Waving Hand Cartoon Vector.svg from Wikimedia Commons by Videoplasty.com, CC-BY-SA 4.0.}}
  \label{fig:man_2d_affine}
  \vspace{0pt}
\end{figure}

\setcounter{figure}{12}
\begin{figure}[t]%
  \includegraphics[width=\linewidth]{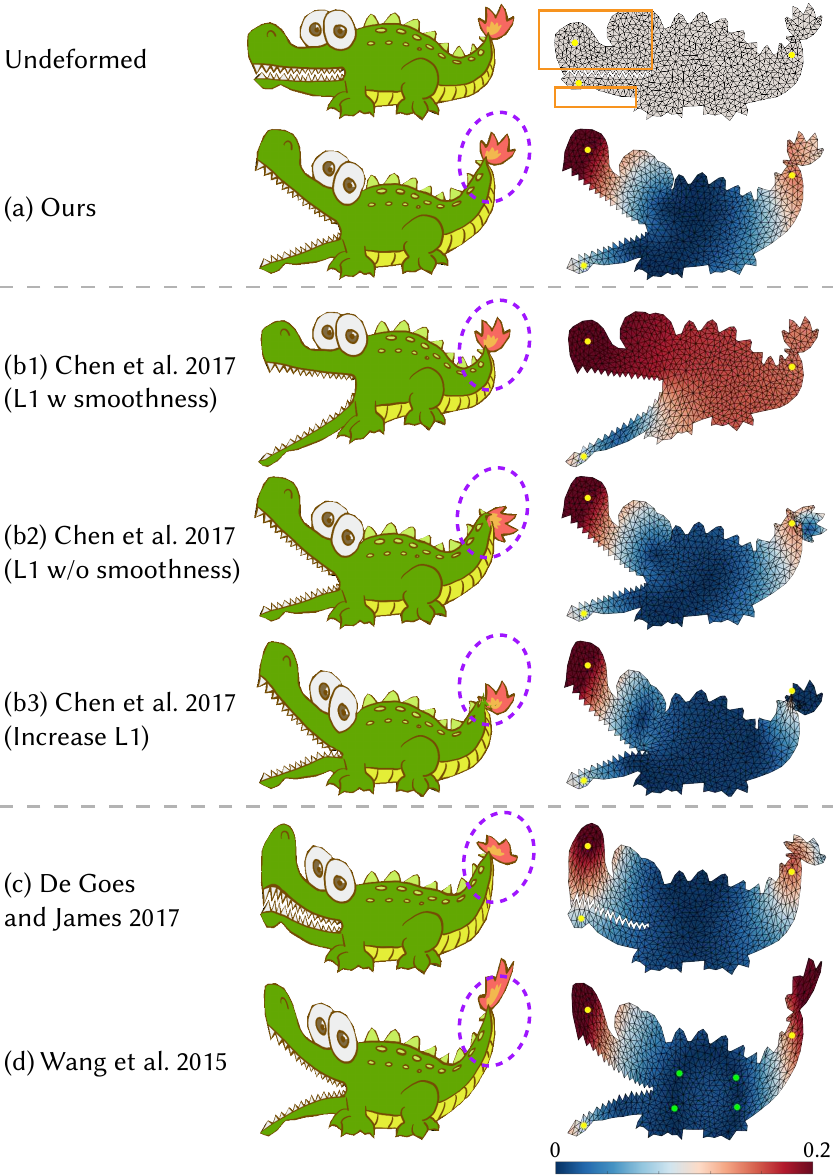}
  \caption{Our local deformation (a) is natural and shape-aware and supports an adaptive ROI without the need of additional handles. 
  Here we highlight all the handles that have been moved in yellow, all the additional fixed handles in green and the affine constraint regions in orange.
  Other local editing alternatives either introduces artifacts (see the flame on the tail in b and d), is unaware of the geometry (see the teeth region in c) or requires additional handles setup (d). 
  \changed{
    More specifically, \cite{chen2017L1ARAP} uses a group lasso penalty, which yields hard to avoid artifacts (b).  
    They therefore mitigate it with a smoothing regularizer, at the expense of fidelity in rotations (b1).  
    Removing the smoothing leads to other artifacts and still has global motion (b2).  
    Further increasing the group lasso penalty removes the global motion but amplifies the artifacts (b3).
  }
  }
  \label{fig:crocodile_bird_vs_all}
\end{figure}

\clearpage %



\section*{Supplementary material}

# Supplementary Material for Local Deformation for Interactive Shape Editing

Honglin Chen
Columbia University
New York City, NY, USA
honglin.chen@columbia.edu

Changxi Zheng
Columbia University
New York City, NY, USA
cxz@cs.columbia.edu

Kevin Wampler
Adobe Research
Seattle, WA, USA
kwampler@adobe.com

## 1 ADDITIONAL RESULTS

We report the runtime of our method and [Chen et al. 2017] on a 2D ARAP local energy and across different mesh resolutions and deformations in Table 1. We also provide the timings and experiment setup of all our 3D examples in Table 2 and the convergence plot in Fig. 1. In addition to the results in Sec. 5, we include an additional example with one or multiple control points in Fig. 2.

Table 1: Timings comparing our ARAP ADMM formulation in Sec.4.1 with standard local-global ARAP and [Chen et al. 2017] for moving the top-right arm of the octocat in Fig.3 of the main text, meshed at three different resolutions. Wall-clock time is measured by averaging over 100 random deformations and uses the (relatively tight) convergence criteria from [Chen et al. 2017]. All timings are generated by single-threaded C++ implementations, written as equivalently as possible between the methods. For an apples-to-apples comparison these timings include the $E_{\text{smooth}}$ energy from [Chen et al. 2017] in the SC-L1 loss and ARAP elastic energies.

|  | ours (w. SC-L1 loss) | ARAP | [Chen et al. 2017] |
|---|---|---|---|
| # triangles | small deformation | | |
| 200 | 0.00087s | 0.00015s | 0.11s |
| 1000 | 0.0055s | 0.0010s | 0.59s |
| 10000 | 0.058s | 0.017s | 2.05s |
| # triangles | large deformation | | |
| 200 | 0.0053s | 0.00032s | 0.022s |
| 1000 | 0.0078s | 0.0018s | 0.17s |
| 10000 | 0.095s | 0.027s | 2.03s |

Authors' addresses: Honglin Chen, Columbia University, New York City, NY, USA, honglin.chen@columbia.edu; Changxi Zheng, Columbia University, New York City, NY, USA, cxz@cs.columbia.edu; Kevin Wampler, Adobe Research, Seattle, WA, USA, kwampler@adobe.com.

## 2 DERIVATION OF PROXIMAL OPERATOR FOR SC-L1 LOSS

For simplicity, we start the derivation with scalar function $f(v) = |v|_{\text{SC-L1}}$. However, the following derivation can be easily extended to vector function $f(\mathbf{x}) = \|\mathbf{x}\|_{\text{SC-L1}}$ by replacing the $|\cdot|$ with $\|\cdot\|$.

For scalar $v$,

$$f(v) = \begin{cases} |v| - \frac{1}{2s}v^2 & |v| < s \\ \frac{1}{2}s & |v| \geq s \end{cases} \quad (1)$$

The gradient of $f(v)$ is

$$f'(v) = \begin{cases} \text{sgn}(v) - \frac{1}{s}v & |v| < s \\ 0 & |v| \geq s \end{cases} \quad (2)$$

The proximal operator of $f(v)$ can be evaluated as

$$\underset{v}{\arg\min} \quad f(v) + \frac{\rho}{2}(v-x)^2 \quad (3)$$

The minimum can be found by setting the derivative to be 0, thus we have

$$0 = f'(v) + \rho(v - x) \quad (4)$$

$$0 = \rho(v-x) + \begin{cases} \text{sgn}(v) - \frac{1}{s}v & |v| < s \\ 0 & |v| \geq s \end{cases} \quad (5)$$

$$0 = \begin{cases} \left(\rho - \frac{1}{s}\right)v + \text{sgn}(v) - \rho x & |v| < s \\ \rho(v - x) & |v| \geq s \end{cases} \quad (6)$$

The location of the minumum $v$ depends on whether or not $\rho > \frac{1}{s}$.

**Case 1:** $\rho \leq \frac{1}{s}$

When $\rho \leq \frac{1}{s}$, then the proximal operator may have up to two local minima—one at the origin and one in the flat region where $|x| \geq s$. We arbitrarily prefer the minima at the origin if it exists, giving:

$$S(x) = \begin{cases} 0 & |x| < s \\ x & |x| \geq s \end{cases} \quad (7)$$

**Case 2:** $\rho \geq \frac{1}{s}$

When $\rho \geq \frac{1}{s}$, then there is only one (global) minimum, so the optimal $v$ is:

$$S(x) = \begin{cases} 0 & |x| \leq \frac{1}{\rho} \\ \frac{\rho s x - s \cdot \text{sgn}(x)}{\rho s - 1} & \frac{1}{\rho} < |x| < s \\ x & |x| \geq s \end{cases} \quad (8)$$

for a vector $x$ this can be further simplied to be

$$S(x) = \begin{cases} \left(\frac{\rho s - s/\|x\|_2}{\rho s - 1}\right)_+ x, & \text{if } \|x\|_2 \leq s \\ x, & \text{otherwise} \end{cases} \quad (9)$$

Table 2: Performance of and parameters for our algorithm on all the 3D examples. All wall-clock timings are reported in seconds, physical parameters are reported with appropriate units. Timing experiments are measured with 8-thread parallelization and PARDISO support [Alappat et al. 2020; Bollhöfer et al. 2019, 2020]. |V| is the number of vertices, |T| is the number of elements. $\rho$ is the ADMM penalty parameter and $s$ is the SC-L1 loss threshold. Local 1, Local 2 and Global is the total runtime for the local step 1, the local step 2 and the global step respectively. Total is the total runtime for our ADMM algorithm. Residual is the final residuals of our energy, defined as $|E_{\text{curr}} - E_{\text{prev}}|/(E_{\text{curr}} + 1.0)$, where $E$ is the total energy and adding 1.0 is to avoid dividing by 0. We use $w = 1e^4$ as the weight of the SC-L1 loss for all the 3D examples. The Young's modulus is $1e^7$ Pa and the poisson's ratio is 0.45 for all our Neo-Hookean examples. We use different $\rho$ for ARAP and Neo-Hookean energy to handle the different magnitude of the energies due to the physical parameters in Neo-Hookean energy. In general we set $s$ to be 0.01 times the max extent of the mesh's bounding box, but other values for $s$ also work well. For interactive frame rates, we generate all the results using a fixed 500 ADMM iterations or when its objective residual reaches $1e^{-11}$. As shown by Fig. 1, all of examples converged within 500 iterations. For the bar example in Fig.7 and Fig.13 of the main text, we report its average runtime.

| Example | |V| | |T| | Material | $\rho$ | $s$ | Residual | Local 1(s) | Local 2(s) | Global(s) | Total(s) |
|---|---|---|---|---|---|---|---|---|---|---|
| **Cloth 1(Fig. 6)** | 14,641 | 28,800 | Cloth | $5e^3$ | 0.4 | $3.93e^{-5}$ | 1.58 | 0.04 | 1.76 | 3.38 |
| **Cloth 2(Fig. 6)** | 14,641 | 28,800 | Cloth | $5e^3$ | 0.4 | $2.09e^{-5}$ | 1.52 | 0.03 | 1.74 | 3.29 |
| **Cloth 1(Suppl Fig.3)** | 14,641 | 28,800 | Cloth | $3e^6$ | 0.01 | $5.04e^{-5}$ | 1.95 | 0.05 | 1.88 | 3.89 |
| **Cloth 2(Suppl Fig.3)** | 14,641 | 28,800 | Cloth | $3e^6$ | 0.01 | $3.54e^{-7}$ | 1.60 | 0.06 | 1.71 | 3.37 |
| **Cloth(Fig. 1)** | 46,977 | 92,928 | Cloth | $3e^6$ | 0.01 | $3.26e^{-4}$ | 6.06 | 0.10 | 5.41 | 11.57 |
| **Pig(Fig. 1)** | 10,416 | 43,749 | NH | $5e^5$ | 0.01 | $3.58e^{-3}$ | 5.85 | 0.03 | 1.61 | 7.49 |
| **Pig 1(Fig. 11)** | 10,416 | 43,749 | ARAP | $1e^{-1}$ | 0.01 | $3.84e^{-7}$ | 2.40 | 0.03 | 1.59 | 4.03 |
| **Pig 2(Fig. 11)** | 10,416 | 43,749 | NH | $5e^5$ | 0.01 | $3.94e^{-5}$ | 5.56 | 0.03 | 1.58 | 7.18 |
| **Cat(Fig. 1)** | 9,078 | 36,115 | NH | $5e^3$ | 0.01 | $7.81e^{-6}$ | 4.24 | 0.02 | 1.30 | 5.56 |
| **Robot(Fig. 1)** | 6,601 | 26,024 | NH | $1e^3$ | 0.04 | $6.88e^{-4}$ | 3.91 | 0.04 | 0.86 | 4.81 |
| **Cube(Fig. 1)** | 4,913 | 20,480 | NH | $1e^5$ | 1.0 | $1.14e^{-5}$ | 2.77 | 0.01 | 2.17 | 4.95 |
| **Apple(Fig. 1)** | 1,088 | 4,072 | NH | $1e^3$ | 0.01 | $1.69e^{-3}$ | 0.60 | 0.01 | 0.52 | 1.13 |
| **Table(Fig. 1)** | 6,483 | 21,864 | ARAP | $2e^{-4}$ | 0.01 | $3.41e^{-11}$ | 1.46 | 0.04 | 1.03 | 2.53 |
| **Bar(Fig. 8)** | 7,011 | 34,557 | ARAP | $1e^{-1}$ | 0.1 | $1.28e^{-4}$ | 2.89 | 0.08 | 2.69 | 5.65 |
| **Bar(Fig. 5)** | 7,011 | 34,557 | NH | $5e^5$ | 0.1 | $6.66e^{-5}$ | 5.74 | 0.03 | 1.63 | 7.41 |

## 3 DERIVATION OF LOCAL ARAP ENERGY

With as-rigid-as-possible (ARAP) energy [Sorkine and Alexa 2007] as our elastic energy, the total energy for our local deformation is as follows:

$$\underset{\mathbf{V},\{\mathbf{R}_i\}}{\text{minimize}} \sum_{i \in V} \sum_{j \in \mathcal{N}(j)} \underbrace{\frac{w_{ij}}{2} \|\mathbf{R}_i \widetilde{\mathbf{d}}_{ij} - \mathbf{d}_{ij}\|_2^2}_{\text{ARAP}} + \underbrace{w a_i \|\mathbf{V}_i - \widetilde{\mathbf{V}}_i\|_{\text{SC-L1}}}_{\text{Localness}}, \quad (10)$$

where $\mathbf{R}_i$ is a $d \times d$ rotation matrix, $w_{ij}$ is the cotangent weight, $\mathbf{d}_{ij} = [\mathbf{V}_j - \mathbf{V}_i]^\top$ and $\widetilde{\mathbf{d}}_{ij} = [\widetilde{\mathbf{V}}_j - \widetilde{\mathbf{V}}_i]^\top$ are the edge vectors between vertices $i, j$ at the deformed and rest states respectively. $\mathcal{N}(j)$ denotes the neighboring edges of the $i$-th vertex in the style of "spokes-and-rims". Here we use $\mathbf{R}_i$ to denote $\mathbf{X}_i$, since we drive the deformation gradient towards a rotation matrix in ARAP energy.

Our local ARAP deformation energy (Eq. 10) can be rewritten as

$$\underset{\mathbf{V},\{\mathbf{R}_i\}}{\text{minimize}} \sum_{i \in V} \frac{1}{2} \|\mathbf{R}_i \mathbf{D}_i - \widetilde{\mathbf{D}}_i\|_{\mathbf{W}_i}^2 + w a_i \|\mathbf{V}_i - \widetilde{\mathbf{V}}_i\|_{\text{SC-L1}}, \quad (11)$$

where $W_i$ is a $|\mathcal{N}(i)| \times |\mathcal{N}(i)|$ diagonal matrix of cotangent weights, $\widetilde{\mathbf{D}}_i$ and $\mathbf{D}_i$ are $3 \times |\mathcal{N}(i)|$ matrices of "spokes and rims" edge vectors of the $i$-th vertex at the rest and deformed states respectively. $\|\mathbf{X}\|_{\mathbf{W}_i}^2$ denotes $\text{Tr}(\mathbf{X}^\top \mathbf{W}_i \mathbf{X})$.

By setting $\mathbf{Z}_i = \mathbf{V}_i - \widetilde{\mathbf{V}}_i$, we can further rewrite Eq. 11 as

$$\underset{\mathbf{V},\{\mathbf{R}_i\},\mathbf{Z}}{\text{minimize}} \sum_{i \in V} \frac{1}{2} \|\mathbf{R}_i \mathbf{D}_i - \widetilde{\mathbf{D}}_i\|_{\mathbf{W}_i}^2 + w a_i \|\mathbf{Z}_i\|_{\text{SC-L1}}, \quad (12a)$$

$$\text{s.t.} \quad \mathbf{Z}_i = \mathbf{V}_i - \widetilde{\mathbf{V}}_i, \quad \forall i \quad (12b)$$

The ADMM update for the above minimization problem is as follows:

$$\mathbf{R}_i^{k+1} \leftarrow \underset{\mathbf{R}_i \in SO(3)}{\arg\min} \frac{1}{2} \|\mathbf{R}_i \mathbf{D}_i - \widetilde{\mathbf{D}}_i\|_{\mathbf{W}_i}^2 \quad (13a)$$

$$\mathbf{Z}_i^{k+1} \leftarrow \underset{\mathbf{Z}_i}{\arg\min} \; w a_i \|\mathbf{Z}_i\|_{\text{SC-L1}} + \frac{\rho}{2} \|\mathbf{V}_i^{k+1} - \widetilde{\mathbf{V}}_i - \mathbf{Z}_i + \mathbf{U}_i^k\|_2^2 \quad (13b)$$

$$\mathbf{V}^{k+1} \leftarrow \underset{\mathbf{V}}{\arg\min} \; \mathbf{W}(\mathbf{V}^\top \mathbf{L} \mathbf{V} - \mathbf{B}^\top \mathbf{V}) + \frac{\rho}{2} \|\mathbf{V} - \widetilde{\mathbf{V}} - \mathbf{Z}^k + \mathbf{U}^k\|_2^2 \quad (13c)$$

$$\mathbf{U}_i^{k+1} \leftarrow \mathbf{U}_i^k + \mathbf{V}_i^{k+1} - \widetilde{\mathbf{V}}_i - \mathbf{Z}_i^{k+1} \quad (13d)$$

Here $\rho$ is a fixed penalty parameter.

**Updating R**

Local step 1 (Eq. 13a) is an instance of the Orthogonal Procrustes problem, which can be solved in the same way as the rotation fitting

**ALGORITHM 1:** Three-block ADMM for ARAP energy

**Input:** A triangle or tetrahedral mesh $\widetilde{V}, T$
**Output:** Deformed vertex positions $V$
$V \leftarrow \widetilde{V}$
**while** *not converged* **do**
$\quad R_i \leftarrow local\_step\_X(V, \widetilde{V})$ ▷ local step 1
$\quad Z_i \leftarrow local\_step\_Z(V, \widetilde{V})$ ▷ local step 2
$\quad V \leftarrow global\_step(X_i, Z_i, \widetilde{V})$ ▷ global step
$\quad U_i \leftarrow dual\_update(Z_i, V_i, \widetilde{V}_i)$ ▷ dual update 1
**end**

step in [Sorkine and Alexa 2007]:

$$R_i^{k+1} \leftarrow \arg\max_{R_i \in SO(3)} \text{Tr}(R_i M_i) \quad (14)$$

$$M_i = D_i \widetilde{D}_i^\top \quad (15)$$

One can derive the optimal $R_i$ from the singular value decomposition of $M_i = \mathcal{U}_i \Sigma_i \mathcal{V}_i^\top$:

$$R_i^{k+1} \leftarrow \mathcal{V}_i \mathcal{U}_i^\top \quad (16)$$

**Updating Z**

Local step 2 (Eq. 13b) can be solved with a shrinkage step (see derivation in App. 2):

$$Z_i^{k+1} \leftarrow \mathcal{S}_{wa_i}^k \left(V_i - \widetilde{V}_i + U_i\right) \quad (17)$$

$$\mathcal{S}_{wa_i}(x) = \begin{cases} \left(\frac{\rho s - wa_i s/\|x\|_2}{\rho s - wa_i}\right)_+ x, & \text{if } \|x\|_2 \le s \\ x, & \text{otherwise} \end{cases} \quad (18)$$

where $\rho$ is set to satisfy $\rho > \frac{\max(wa_i)}{s}$ (see App. 2).

**Updating V**

The global step (Eq. 13c) can be achieved by solving a linear system:

$$(L + \rho I)V = B + \rho(\widetilde{V} + Z^k - U^k), \quad (19)$$

which can be precomputed for fixed $\rho$.

## 4 DERIVATION OF LOCAL NEO-HOOKEAN ENERGY

Our local deformation scheme can be further extended to physics-based elasticity energies, e.g., Neo-Hookean energy. Using the Neo-Hookean energy as our elasticity energy, the optimization problem can be written as follows:

$$\min_{V, \{X_j\}, \{Z_i\}} \sum_{j \in T} \underbrace{E_{\text{nh}}(X_j)}_{\text{Neo-Hookean}} + \sum_{i \in V} \underbrace{wa_i \|V_i - \widetilde{V}_i\|_{\text{SC-L1}}}_{\text{Localness}}, \quad (20a)$$

$$\text{s.t.} \quad X_j = \text{sym}(D_j V), \forall j, \quad (20b)$$

where $T$ denotes all the elements.

By introducing $Z_i = V_i - \widetilde{V}_i$, Eq. 20 can be further rewritten into:

$$\min_{V, \{X_j\}, \{Z_i\}} \sum_{j \in T} E_{\text{nh}}(X_j) + \sum_{i \in V} wa_i \|Z_i\|_{\text{SC-L1}}, \quad (21a)$$

$$\text{s.t.} \quad X_j = \text{sym}(D_j V), \forall j, \quad (21b)$$

$$Z_i = V_i - \widetilde{V}_i, \quad \forall i \quad (21c)$$

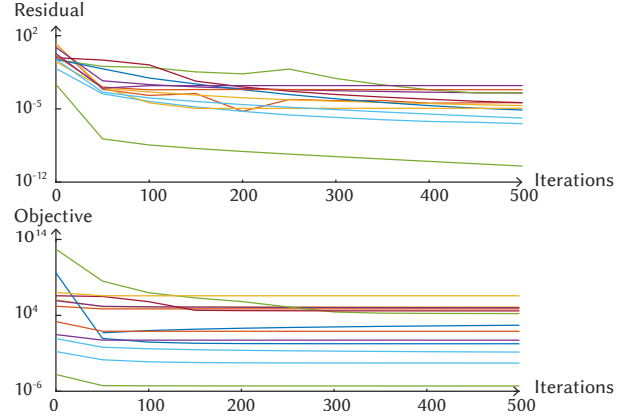

Figure 1: Convergence plot. We demonstrate the convergence of our ADMM by plotting residuals ($|E_{\text{curr}} - E_{\text{prev}}|/(E_{\text{curr}} + 1.0)$) and objective values of all our examples in Table 2 in the main text over the ADMM iterations. As can be seen from the plot, all our 3D examples converge within 500 ADMM iterations.

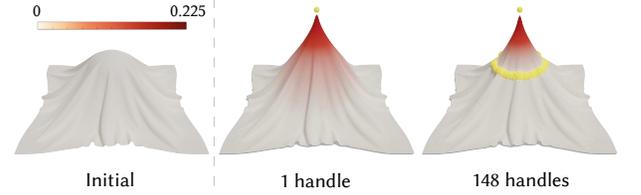

Figure 2: Our local deformer allows the user to use just a single control point, without the need to "pin down" the mesh. It also supports adding more control points to achieve more detailed control.

The augmented Lagrangian for this problem (Eq. 21) is:

$$L(V; \{X_i\}, \{W_i\}; \{Z_i\}, \{U_i\}) = \sum_{j \in T} E_{\text{nh}}(X_j) + \sum_{i \in V} wa_i \|Z_i\|_{\text{SC-L1}}$$
$$+ \sum_{j \in T} \frac{\gamma}{2} \|\text{sym}(D_j V) - X_j + W_j\|_2^2 + \sum_{i \in V} \frac{\rho}{2} \|V - \widetilde{V}_i - Z_i + U_i\|_2^2. \quad (22)$$

Therefore, the ADMM update for the above minimization problem is as follows:

$$X_j^{k+1} \leftarrow \arg\min_{X_j} E_{\text{nh}}(X_j) + \frac{\gamma}{2} \|\text{sym}(D_j V) - X_j + W_j\|_2^2 \quad (23a)$$

$$Z_i^{k+1} \leftarrow \arg\min_{Z_i} wa_i \|Z_i\|_{\text{SC-L1}} + \frac{\rho}{2} \|V_i^{k+1} - \widetilde{V}_i - Z_i + U_i^k\|_2^2 \quad (23b)$$

$$V^{k+1} \leftarrow \arg\min_V \sum_{j \in T} \frac{\gamma}{2} \|\text{sym}(D_j V) - X_j + W_j\|_2^2 + \sum_{i \in V} \frac{\rho}{2} \|V - \widetilde{V} - Z^k + U^k\|_2^2 \quad (23c)$$

$$W_j^{k+1} \leftarrow W_j^k + \text{sym}(D_j V) - X_j \quad (23d)$$

$$U_i^{k+1} \leftarrow U_i^k + V_i^{k+1} - \widetilde{V}_i - Z_i^{k+1} \quad (23e)$$

Here $\rho$ and $\gamma$ are fixed penalty parameters.

Compared with the ADMM steps for the local ARAP energy (Eq. 13), the local step 2 ($\arg\min_{Z_i}$) and the global step ($\arg\min_V$)

**ALGORITHM 2:** Three-block ADMM for Neo-Hookean energy

**Input:** A triangle or tetrahedral mesh $\widetilde{\mathbf{V}}, \mathbf{T}$
**Output:** Deformed vertex positions $\mathbf{V}$
$\mathbf{V} \leftarrow \widetilde{\mathbf{V}}$
**while** *not converged* **do**
  $\quad \mathbf{X}_i \leftarrow local\_step\_X(\mathbf{V}, \widetilde{\mathbf{V}})$ ▷ local step 1
  $\quad \mathbf{Z}_i \leftarrow local\_step\_Z(\mathbf{V}, \widetilde{\mathbf{V}})$ ▷ local step 2
  $\quad \mathbf{V} \leftarrow global\_step(\mathbf{X}_i, \mathbf{Z}_i, \widetilde{\mathbf{V}})$ ▷ global step
  $\quad \mathbf{U}_i \leftarrow dual\_update(\mathbf{Z}_i, \mathbf{V}_i, \widetilde{\mathbf{V}}_i)$ ▷ dual update 1
  $\quad \mathbf{W}_i \leftarrow dual\_update(\mathbf{X}_i, \mathbf{V}_i, \widetilde{\mathbf{V}}_i)$ ▷ dual update 2
**end**

for the local Neo-Hookean energy can be solved in the same way as the local ARAP energy (see App. 3).

**Updating X**

Local step 1 (Eq. 23a) can be solved by performing the energy minimization on the singular values of $\mathrm{sym}(\mathbf{D}_j \mathbf{V}) + \mathbf{W}_j$. This is a proximal operator of $E_{\mathrm{nh}}$ at rotation-invariant $\mathrm{sym}(\mathbf{D}_j \mathbf{V}) + \mathbf{W}_j$.

Let us denote the proximal operator of $E_{\mathrm{nh}}$ and the singular value decomposition of $\mathrm{sym}(\mathbf{D}_j \mathbf{V}) + \mathbf{W}_j$ as:

$$\mathrm{prox}_{E_{\mathrm{nh}}}(\mathbf{X}_j) = E_{\mathrm{nh}}(\mathbf{X}_j) + \frac{\gamma}{2} \|\mathrm{sym}(\mathbf{D}_j \mathbf{V}) + \mathbf{W}_j - \mathbf{X}_j\|_2^2 \quad (24)$$

$$\mathrm{sym}(\mathbf{D}_j \mathbf{V}) + \mathbf{W}_j = \mathcal{U}_j \Sigma_j \mathcal{V}_j^\top \quad (25)$$

As shown by [Brown and Narain 2021], we can compute the optimal $\mathbf{X}_j$ as:

$$\Sigma_j^{k+1} \leftarrow \arg\min_{\Sigma_j} \mathrm{prox}_{E_{\mathrm{nh}}}(\mathcal{U}_j \Sigma_j \mathcal{V}_j^\top) \quad (26)$$

$$\mathbf{X}_j \leftarrow \mathcal{U}_j \Sigma_j^{k+1} \mathcal{V}_j^\top \quad (27)$$

Specifically, one can compute the SVD of $\mathrm{sym}(\mathbf{D}_j \mathbf{V}) + \mathbf{W}_j$ and perform the minimization of $\mathrm{prox}_{E_{\mathrm{nh}}}$ only on its singular values, while keeping singular vectors unchanged. The above optimization of the singular values $\Sigma_j^{k+1}$ can be performed using an L-BFGS solver.

## 5 OPTIMAL SCALING $s_k$ FOR $E_{\mathrm{ACAP}}$

Adopting the as-conformal-as-possible energy $E_{\mathrm{ACAP}}$ as our elastic energy requires solving an instance of isotropic orthogonal Procrustes problem in the local step 1 (updating $\mathbf{R}_k$). Following the derivation in [Schönemann and Carroll 1970], one can compute its analytical solution as follows: The optimal rotation $\mathbf{R}_k$ can be computed the same way as the local step 1 (updating $\mathbf{R}_k$) for the ARAP energy, and the optimal scaling $s_k$ can be computed analytically as

$$s_k = \frac{\mathrm{Tr}\left(\mathbf{W}_k \widetilde{\mathbf{D}}_k^\top \mathbf{R}_k \mathbf{D}_k\right)}{\mathrm{Tr}\left(\mathbf{W}_k \mathbf{D}_k^\top \mathbf{D}_k\right)} \quad (28)$$

When assembling the matrices $\mathbf{B}$ for the global step, one needs to replace $\mathbf{R}_k$ with $s_k \mathbf{R}_k$.

## 6 COMPARISON WITH OTHER SC-L1 LOSS ALTERNATIVES

Our SC-L1 loss is designed with the following desirable properties in mind:

(1) Behaves like a $\ell_1$ loss near zero.
(2) Is constant far from zero.
(3) Admits an efficient proximal shrinkage operator free of local minima.

We will consider each of these properties in turn, examining the motivation and comparing our loss to other alternatives.

*(1) Behaves like a $\ell_1$ loss near zero.* This property is what gives our method the ability to localize deformation. The non-differentiable $\ell_1$-like "vertex" of the SC-L1 loss at the origin is critical in this, and we refer the reader to [Hastie et al. 2015] Sec. 2.2 for more information as to why. Using a loss which is smooth at the origin will invariably result in small amounts of global motion, with the degree of global motion being roughly related to the magnitude of the function's Hessian at the origin, so the smoother the loss, the more global motion should be expected. Examples of such smooth losses are the smoothed-$\ell_1$ or Huber loss [Huber 1964] or smoothed-$\ell_0$ losses [Mohimani et al. 2007; Xiang et al. 2022], and we avoid these losses due to their inability to eliminate global deformations.

*(2) Constant far from zero.* This property has the effect of disabling the locality-inducing force for parts of a deformed shape which have moved beyond a certain radius, and serves to remove artifacts resulting from parts of a deformed shape being pulled toward a rest position which is now far away. Fig.3(b) and Fig.10(b) in the main text illustrate the types of artifacts which can be observed using a loss without this property.

*(3) Efficient proximal shrinkage operator free of local minima.* It is convenient for the loss' proximal shrinkage operator to be uniquely defined. Recalling that this operator is defined as the minimizer(s) of Eq. 3, the condition that it be uniquely defined is equivalent to requiring that Eq. 3 has a unique minimum. A sufficient condition for this is that $f(v) + \frac{\rho}{2}(v-x)^2$ is strictly convex, or equivalently that $\rho > -\frac{d^2}{dv^2} f(v)$. We thus prefer a loss function which has a bounded negative second derivative, and the lower this bound the lower the value of $\rho$ which can be used to convexify it. Within distance $s$ from the origin (excluding the origin itself), our loss has a constant negative value for $\frac{d^2}{dv^2} f(v)$, making it in some sense the minimal function satisfying the three desired properties. Other losses are of course possible, but would require a higher value of $\rho$ for the same radius of effect.

*Alternative losses.* A natural question is whether there might be other locality-inducing losses which could be used in place of our SC-L1 loss, perhaps even a generic spline-based loss which could be shaped in detail. Although we have not directly explored this, we expect that there would be many possible options provided they exactly or sufficiently approximately satisfy the three desired properties outlined above. That said, in our experiments we have found that it's normally useful to set $s$ to be relatively small to limit artifacts. In this situation the specific shape of the loss is not particularly important since it only has an impact within a visually small region, and our simple formulation seems to work well.

## REFERENCES
Christie Alappat, Achim Basermann, Alan R. Bishop, Holger Fehske, Georg Hager, Olaf Schenk, Jonas Thies, and Gerhard Wellein. 2020. A Recursive Algebraic Coloring Technique for Hardware-Efficient Symmetric Sparse Matrix-Vector Multiplication.



\end{document}